\documentclass[aps,pra,10pt,twocolumn,letterpaper,superscriptaddress,showpacs,floatfix]{revtex4-1}
\usepackage{subfig}
\usepackage{color}
\usepackage{graphicx}
\usepackage{amsmath}
\usepackage{amssymb,amsthm}
\usepackage{hyperref}
\usepackage{mathtools}

\usepackage[applemac]{inputenc}

\hypersetup{colorlinks=true,linkcolor=blue,citecolor=blue,urlcolor=blue}

\begin{document}
	
\title{Estimation of pure states using three measurement bases}
	
\author{L. Zambrano}
\email[corresponding author:]{leozambrano@udec.cl}	
\affiliation{Instituto Milenio de Investigaci\'on en \'Optica, Universidad de Concepci\'on, Concepci\'on, Chile}
\affiliation{Facultad de Ciencias F\'isicas y Matem\'aticas, Departamento de F\'isica, Universidad de Concepci\'on, Concepci\'on, Chile}

\author{L. Pereira}
\affiliation{Instituto Milenio de Investigaci\'on en \'Optica, Universidad de Concepci\'on, Concepci\'on, Chile}
\affiliation{Facultad de Ciencias F\'isicas y Matem\'aticas, Departamento de F\'isica, Universidad de Concepci\'on, Concepci\'on, Chile}

\author{D. Mart\'inez}
\affiliation{Instituto Milenio de Investigaci\'on en \'Optica, Universidad de Concepci\'on, Concepci\'on, Chile}
\affiliation{Facultad de Ciencias F\'isicas y Matem\'aticas, Departamento de F\'isica, Universidad de Concepci\'on, Concepci\'on, Chile}

\author{G. Ca\~nas}
\affiliation{Instituto Milenio de Investigaci\'on en \'Optica, Universidad de Concepci\'on, Concepci\'on, Chile}
\affiliation{Departamento de F\'isica, Universidad del B\'io-B\'io, Collao 1202, Casilla 5C, Concepci'on, Chile}

\author{G. Lima}
\affiliation{Instituto Milenio de Investigaci\'on en \'Optica, Universidad de Concepci\'on, Concepci\'on, Chile}
\affiliation{Facultad de Ciencias F\'isicas y Matem\'aticas, Departamento de F\'isica, Universidad de Concepci\'on, Concepci\'on, Chile}

\author{A. Delgado}
\affiliation{Instituto Milenio de Investigaci\'on en \'Optica, Universidad de Concepci\'on, Concepci\'on, Chile}
\affiliation{Facultad de Ciencias F\'isicas y Matem\'aticas, Departamento de F\'isica, Universidad de Concepci\'on, Concepci\'on, Chile}
	
\date{\today}
	
\begin{abstract}
We introduce a new method to estimate unknown pure $d$-dimensional quantum states using the probability distributions associated with only three measurement bases. Measurement results of $2d$ projectors are employed to generate a set of $2^{d-1}$ possible states, the likelihood of which is evaluated using the measurement results of the $d$ remaining projectors. The state with the highest likelihood is the estimate of the unknown state. The method estimates all pure states but a null-measure set. The viability of the protocol is experimentally demonstrated using two different and complementary high-dimensional quantum information platforms. First, by exploring the photonic path-encoding strategy, we validate the method on a single 8-dimensional quantum system. Then, we resort to the five superconducting qubit IBM quantum processor to demonstrate the high performance of the method in the multipartite scenario.
\end{abstract}

\maketitle

\section{Introduction}
		
The estimation of unknown quantum states of high dimension and the assessment of quantum processes and devices has proven to be a remarkably difficult task from an experimental \cite{Haeffner,Monz,Wang,Gong} and theoretical \cite{Silva,Smolin,Shang} point of view. The estimation of unknown quantum states requires the acquisition of information by means of measurements \cite{SQT1,SQT2} and its subsequent post-processing \cite{MLE1,MLE2,MLE3,BAYES}. For a single $d$-dimensional quantum system, the minimal total number of measurement outcomes required to estimate an unknown state is $d^2$. Various estimation methods employ a number of measurement outcomes that is equal to or very similar to $d^2$. Symmetric informationally complete positive operator-valued measure (SIC-POVM) \cite{SIC1,SIC2,SIC3,SIC4,USD,SIC5,SIC6,SIC7} are generalized measurements that allow for estimating quantum states with exactly $d^2$ measurement outcomes. Mutually unbiased bases (MUBs) \cite{MUB1,MUB2,MUB3,MUB4,MUB5,MUB6,MUB7} estimate quantum states with $d^2+d$ measurement outcomes. The existence of SIC-POVMs and MUBs has been proven in restricted sets of dimensions. For this reason alternative schemes have been proposed \cite{ADAPTIVE,USDQT,EQT,MS}, which require in the order of $d^2$ measurement outcomes. In the case of a multipartite system formed by  $n$ $d$-dimensional systems, the total number of measurement outcomes becomes $d^{2n}$. In this way, the total number of measurement results grows exponentially with the number of parties, which increases the experimental complexity of the data acquisition process, as well as the computational cost of the optimization problem associated with subsequent data processing.

In order to make the problem tractable, the use of a priori information has been considered. Thereby, the estimation focuses in a restricted set of states, which allows reducing the number of measurements \cite{MPS,CS,Toth,Ahn}. Recently, it has been shown \cite{5B} that a set of four fixed observables are sufficient to estimate pure quantum states up to a statistically unlikely null-measure set. This set is formed by pure states that in the canonical basis have two or more nonconsecutive vanishing coefficients. The addition of a fifth observable, which is diagonal in the canonical basis,  helps determine whether or not a given state belongs to this null-measure set. If this is the case, then the remaining four observables can be adapted to a lower dimensional subspace. This result is independent of the underlying dimension of the Hilbert space for $d>4$. In this way, any pure quantum state can be estimated with a total of $5d$ projective measurements, at most. This estimation procedure involves a simple post-processing stage and allows the purity assumption to be certified directly from the measurement results. Lately, it has been shown \cite{5BFIJAS} that adaptivity it is not necessary. Five fixed observables estimate all pure quantum states in any dimension, at the expense of a more convoluted construction of the observables and a much more complex post-processing stage.
	
Here, we study the estimation of pure quantum states by means of three fixed observables only, that is, with a total of $3d$ projective measurements. In particular, we show that $2d$ projective measurements generate a finite set $\Omega$ of $2^{d-1}$ pure states. Half of the rank-1 projectors comes from an observable, while the remaining rank-1 projectors correspond to half of the projectors of each of the remaining two observables. The estimate for the unknown state is given by the state in $\Omega$ with the highest likelihood, which is evaluated with the measurement results of the remaining $d$ rank-1 projective measurements. Thus, the costly procedure of optimizing the probability is not necessary. The present method estimates all pure quantum states but a null-measure set. We also consider the role of finite statistics effects and show that for moderate ensemble sizes the present estimation method provides results with an accuracy comparable to that achieved by the 5-bases based pure-state quantum tomographic method.

We also demonstrate the experimental feasibility of our estimation method by employing two different and complementary high-dimensional quantum information platforms. First, we estimate the state of an 8-dimensional quantum system that is encoded in the linear transverse momentum of single photons transmitted through diffractive apertures addressed into spatial light modulators \cite{Neves05,Neves07,Glima08,Glima09}. This platform can attain high fidelities for preparing and measuring high-dimensional quantum states \cite{QRAC1024}, and therefore its use allows one to proper address the performance of new methods for quantum state reconstruction in higher dimensions. In this case, we achieve a remarkable fidelity of $98.5\%$ between the reconstructed state and the prepared one.  Then, we study the method in a mutlti-partite scenario and apply it to estimate a two-qubit state generated on the IBM Quantum Experience 5-qubit superconducting quantum processor ``ibmq-ourense". In this case, we are also able to achieve a high fidelity of $96.5\%$. These results highlight the versatility and high-performance of the protocol, indicating that it can be a valuable tool supporting the development of future quantum technologies dealing with more complex quantum systems \cite{Guix_2019}.

\section{Review on the 5-bases based pure-state estimation method}
	
The 5-bases based pure-state quantum tomographic method (5BB-QT) \cite{5B} employs projective measurements onto the canonical basis ${\cal B}_0=\{|i\rangle\}$ (with $i=0,\dots,d-1$) and the bases
\begin{eqnarray}
{\cal B}_1&=&\left\{|\varphi^\nu_{\pm}\rangle_1=\frac{1}{\sqrt{2}}(|2\nu\rangle\pm|2\nu+1\rangle)\right\},
\nonumber\\
{\cal B}_2&=&\left\{|\tilde\varphi^\nu_{\pm}\rangle_2=\frac{1}{\sqrt{2}}(|2\nu\rangle\pm i|2\nu+1\rangle)\right\},
\nonumber\\
{\cal B}_3&=&\left\{|\varphi^\nu_{\pm}\rangle_3=\frac{1}{\sqrt{2}}(|2\nu+1\rangle\pm|2\nu+2\rangle)\right\},
\nonumber\\
{\cal B}_4&=&\left\{|\tilde\varphi^\nu_{\pm}\rangle_4=\frac{1}{\sqrt{2}}(|2\nu+1\rangle\pm i|2\nu+2\rangle)\right\},
\end{eqnarray}
where $\nu\in[0,(d-2)/2]$. Operations with labels are carried out modulo $d$. In the case of odd dimensions, the integer part of $(d-2)/2$ is considered and every basis is completed with the state $|d\rangle$. The 5BB-QT method estimates almost any pure state $| \psi \rangle = \sum_{k=0}^{d-1} c_k | k \rangle $ in any dimension $d$ using the set of probability distributions generated by projections on the bases ${\cal B}_i$ (with $i=1,\dots,4$). The states that cannot be estimated have at least two nonconsecutive vanishing coefficients. In this case, the system of equations to be solved has infinite solutions. In order to avoid this problem, a fifth basis, the canonical one, is introduced. This is the first basis to be measured, and its only purpose it is to detect the states that cannot be estimated. If this is the case, the method is adapted by reducing the effective dimension of the estimated state and the bases.

We define  $p^{(k)}_{\pm}$ with $k$ even (odd) as  $p^{(k)}_{\pm} = |\langle \varphi^k_{\pm} |\psi\rangle|^2$ with $|\varphi^k_{\pm}\rangle$ in basis ${\cal B}_1$ (${\cal B}_3$) and $\tilde p^{(k)}_{\pm}$ with $k$ even (odd) as $\tilde{p}^{(k)}_{\pm} = |\langle \tilde{\varphi}^k_{\pm} |\psi\rangle|^2$  with $|\tilde{\varphi}^k_{\pm}\rangle$ in basis ${\cal B}_2$ (${\cal B}_4$). These quantities correspond to transition probabilities from the unknown state toward the states in the bases ${\cal B}_i$ and can be experimentally measured. In order to estimate the unknown pure state, the 5BB-QT method employs the set of $d$ equations 	
\begin{equation}
2 c_k c_{k+1}^* = \Lambda_k,
\label{5BBQTRULE}
\end{equation}
with
\begin{equation}
\Lambda_k =(p^{(k)}_+-p^{(k)}_-)+i(\tilde p^{(k)}_+-\tilde p^{(k)}_-),
\label{LAMBDAS}
\end{equation}
for $k=0,\dots,d-1$. This set of equations can be iteratively solved for the complex probability amplitudes $c_k$ that characterize the unknown state.

\section{The 3-bases based pure-state estimation method}
We can modify the bases (1) in the following way:
\begin{eqnarray}
{\cal B}_1&=&\left\{|\varphi^\nu_{\pm}\rangle_1=a |2\nu\rangle\pm b|2\nu+1\rangle\right\},
\nonumber\\
{\cal B}_2&=&\left\{|\tilde\varphi^\nu_{\pm}\rangle_2=a|2\nu\rangle\pm ib|2\nu+1\rangle\right\},
\nonumber\\
{\cal B}_3&=&\left\{|\varphi^\nu_{\pm}\rangle_3=a|2\nu+1\rangle\pm b|2\nu+2\rangle\right\},
\nonumber\\
{\cal B}_4&=&\left\{|\tilde\varphi^\nu_{\pm}\rangle_4=a|2\nu+1\rangle\pm ib|2\nu+2\rangle\right\},
\end{eqnarray}
with $|a|^2 + |b|^2 = 1$. In this case Eq.\thinspace(\ref{5BBQTRULE}) is still valid. Equation\thinspace(\ref{LAMBDAS}) becomes
\begin{equation}
	\Lambda_k =\frac{(p^{(k)}_+-p^{(k)}_-)+i(\tilde p^{(k)}_+-\tilde p^{(k)}_-)}{ab}.
	\label{NEWLAMBDAS}
\end{equation}
Using this set of equations we can estimate pure states. The quantities $\Lambda_k$ entering in Eq.\thinspace(\ref{NEWLAMBDAS}) can be cast in the form
\begin{eqnarray}
\Lambda_k&=&\left(p^{(k)}_+ - \frac{|ac_k|^2 - |bc_{k+1}|^2}{ab}\right)
\nonumber\\
&&+i\left(\tilde p^{(k)}_+ - \frac{|ac_k|^2- |bc_{k+1}|^2}{ab} \right),
\label{Lambda_k}
\end{eqnarray}
which is now a function of the probabilities $|c_k|^2$ and $|c_{k+1}|^2$. These are obtained from the measurement on the canonical basis ${\cal B}_0$. Equation (\ref{NEWLAMBDAS}) shows that we only need half the projectors of each basis, plus the values obtained from the canonical basis, for unambiguously estimating the unknown quantum state. The other half of the projectors is redundant because it delivers the same information.

Real $\Re(\Lambda_k)$ and imaginary $\Im(\Lambda_k)$ parts of $\Lambda_k$ are not independent. According to Eq.\thinspace(\ref{5BBQTRULE}), they are related through the constraint
\begin{equation}
\Im(\Lambda_k)^2  = 4  |c_k c_{k+1}|^2  - \Re(\Lambda_k)^2,
\label{3BBQTRULE}
\end{equation}
where the right side is determined by transition probabilities toward the states in ${\cal B}_0$ and ${\cal B}_1$ ($k$ even) or ${\cal B}_3$ ($k$ odd). Therefore, Eq.\thinspace(\ref{3BBQTRULE}) allows us to determine the value of $\Lambda_k$ up to a sign without employing the transition probabilities toward the states in the bases ${\cal B}_2$ and ${\cal B}_4$, that is,
\begin{equation}
\Lambda_{k,\pm}=\Re(\Lambda_k)\pm i|\Im(\Lambda_k)|.
\end{equation}
All possible sign combinations in the $d$ coefficients $\Lambda_k$ lead to $2^{d-1}$ different sets $A_j = \{\tilde{\Lambda}_0, \tilde{\Lambda}_1,\dots, \tilde{\Lambda}_{d-2} \}$, with $\tilde{\Lambda}_k=\Lambda_{k,+}$ or $\tilde{\Lambda}_k=\Lambda_{k,-}$.  Let us note that $\Lambda_{d-1}$ is not used since we have assumed that $c_0$ is a real positive number. Solving Eq.\thinspace(\ref{5BBQTRULE}) for each $A_j$ we obtain a set of $2^{d-1}$ states $\{| \psi^{(k)} \rangle\}$ ($k=1,\dots,2^{d-1}$). All states in this set are characterized by the same set $\{|_1\langle\varphi^\nu_{+}|\psi\rangle|^2, |_3\langle\varphi^\nu_{+}|\psi\rangle|^2, |\langle i|\psi\rangle|^2 \}$ of transition probabilities. Thereby, each state $| \psi^{(k)} \rangle$ can be chosen as an estimate of the unknown state $|\psi\rangle$.

In order to lift the ambiguities in the estimation process, the transition probabilities toward states in bases ${\cal B}_2$ and ${\cal B}_4$ can be measured. Instead, we propose to replace the bases ${\cal B}_1$ and ${\cal B}_3$ by the following modified versions
\begin{eqnarray}
{\cal B}'_1&=&\left\{|\varphi^\nu_{+}\rangle_1=a|2\nu\rangle + b|2\nu+1\rangle, | \varphi_j \rangle_1  \right\},
\nonumber\\
{\cal B}'_3&=&\left\{|\varphi^\nu_{+}\rangle_3=a|2\nu+1\rangle + b|2\nu+2\rangle), | \varphi_j \rangle_3\right\}.
\end{eqnarray}
In these new bases we keep half of the bases that allow us to calculate the real part $\Re(\Lambda_k)$ and add $d/2$ states $| \varphi_j \rangle$ to complete the bases. A possible choice for the bases is
\begin{eqnarray}
{\cal B}'_1&=&\left\{|\varphi^\nu_{+}\rangle_1=\frac{1}{\sqrt{2}}(|2\nu\rangle + |2\nu+1\rangle), | \varphi_j \rangle_1  \right\},
\nonumber\\
{\cal B}'_3&=&\left\{|\varphi^\nu_{+}\rangle_3=\frac{1}{\sqrt{2}}(|2\nu+1\rangle + |2\nu+2\rangle), | \varphi_j \rangle_3\right\}.
\label{NEWB1'B2'}
\end{eqnarray}
with the states $| \varphi_j \rangle_1$ and $| \varphi_j \rangle_3$ being
\begin{eqnarray}
| \varphi_j \rangle_1 &=& \frac{1}{\sqrt{2}}\sum_{m=0}^1\sum_{n=0}^{(d-2)/2} (-1)^m\mathcal{F}_{jn}| 2n+m \rangle,
\label{EXAMPLE1}\\
| \varphi_j \rangle_3 &=& \frac{1}{\sqrt{2}}\sum_{m=0}^1\sum_{n=0}^{(d-2)/2}  (-1)^m\mathcal{F}_{jn}| 2n+m+1 \rangle,
\label{EXAMPLE2}
\end{eqnarray}
where $\mathcal{F}_{jk}$ is the transformation
\begin{eqnarray}
\mathcal{F}_{jk}=\frac{1}{\sqrt{d/2}}e^{i\left[\frac{2\pi jk}{d/2}+\phi_k\right]}.
\end{eqnarray}
The phases $\phi_k$ are chosen in such a way that the vectors $|\varphi_j\rangle$ provide information about the imaginary parts $\Im(\Lambda_k)$.

In order to select one of the states $|\psi^{(k)}\rangle$ as the estimate of $|\psi\rangle$ we resort to the transition probabilities $\{|\langle \varphi_j |\psi \rangle |^2\}$. If every element in the set $\{|\psi^{(k)} \rangle\}$ has a different transition probability on at least one of the states $| \varphi_j \rangle $ we can choose the estimate $|\tilde\psi\rangle$ of $|\psi\rangle$ as the state $|\psi^{(k)}\rangle$ whose measurement outcomes reproduce more faithfully the probability transitions $\{|\langle \varphi_j |\psi \rangle |^2\}$, that is, we choose as estimate the state $|\psi^{(k)}\rangle$ with the highest value of the likelihood function
\begin{equation}
|\tilde\psi\rangle={\rm Arg} \max_k \big\{\Pi_j|\langle \varphi_j |\psi^{(k)} \rangle |^{2f_j}\big\},
\end{equation}
where $f_j$ is the experimentally observed frequency of the projection onto the state $|\varphi_j\rangle$. Let us note that here the likelihood $L$ is evaluated to obtain the estimate and not optimized, which helps to reduce the computational cost of the method. In addition, with the 3-bases based pure-state quantum tomographic method (3BB-QT) we have reduced the number of bases required to estimate an unknown pure state from 5 to 3.

As in the case of the 5BB-QT method, the 3BB-QT method also requires to adapt the bases ${\cal B}'_1$ and ${\cal B}'_3$ in certain cases. In particular, when the unknown state has two or more nonconsecutive vanishing coefficients. These are detected by the measurements carried out on the canonical basis, in which case ${\cal B}'_1$ and ${\cal B}'_3$ are adapted to estimate an unknown state belonging to a known lower dimensional subspace. A similar situation arises due to the use of the likelihood. There exist states $|\psi\rangle$ with estimates $|\psi^{(k)}\rangle$ that have the same value of the likelihood. An extreme example for $a = b = 1/\sqrt{2}$ is the state $|\psi\rangle=(1/\sqrt{d})\sum_{i=1}^d|i\rangle$. In this case all states $|\psi^{(k)}\rangle$ have the same likelihood. Another example is the state $|\psi\rangle=(1/2)(|1\rangle+i|1\rangle+|1\rangle+i|1\rangle)$. This leads to two sets of states $|\psi^{(k)}\rangle$, each set contains states with the same likelihood. From the results of several numerical experiments, we conjecture that the states with at least two pairs of equal probability amplitudes lead to states $|\psi^{(k)}\rangle$ with the same value of the likelihood. This is a null measure set. This problem can be mitigated at a great extent by noting that the bases in Eq.\thinspace(\ref{NEWB1'B2'}) are one of many possible choices. The coefficients $a$ and $b$ and the states $| \varphi_j \rangle_1$ and $| \varphi_j \rangle_2$ can be chosen randomly. Thus, when the measurement on the canonical basis reveal two or more pairs of equal transition probabilities, coefficients $a$ and  $b$ and states $|\varphi_j \rangle_1$ and $| \varphi_j \rangle_3$ are randomly chosen and orthogonalized via the Gramm-Schmidt orthogonalization procedure, which generates new bases ${\cal B}'_1$ and ${\cal B}'_3$ such that the states $|\psi^{(k)}\rangle$ have different likelihood values.

\begin{figure}[t]
	\centering
	\includegraphics[width=0.45\textwidth]{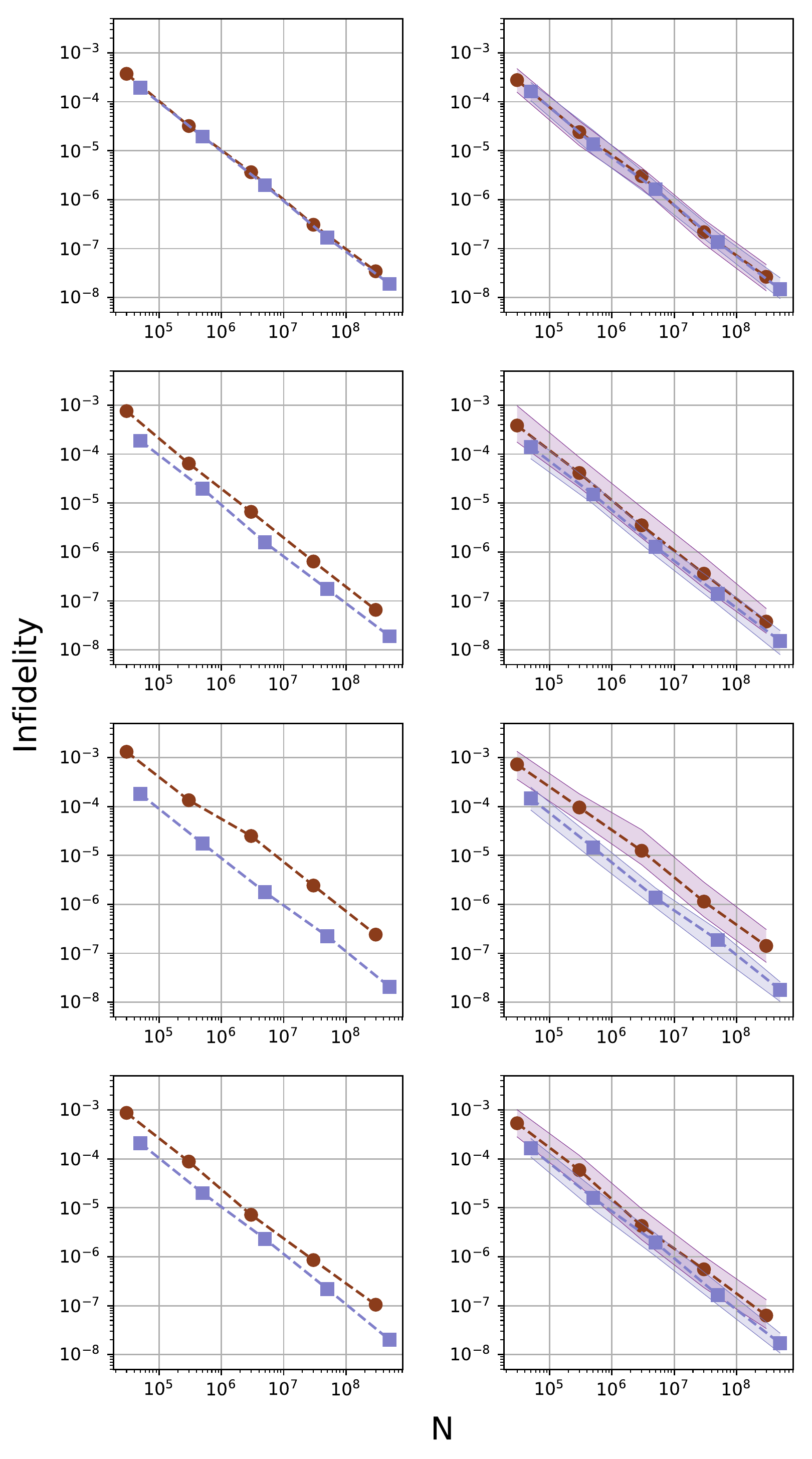}
	\caption{Left (right) column shows in logarithmic scale for both axes the mean (median) infidelity $I(|\psi_j\rangle)$ as a function of the total ensemble size $3N$ obtained via the IC5BB-QT (solid blue squares) and 3BB-QT (solid red dots) methods for four randomly chosen states in $d=4$. Shaded areas represent the corresponding interquartile range.}
\label{Figure1}
\end{figure}

\section{Accuracy of the 3BB-QT method}

In order to test the estimation accuracy achieved by the 3BB-QT method we conduct several numerical experiments. As a figure of merit for the accuracy of the estimation process, we employ the infidelity $I(|\psi\rangle, |\tilde\psi\rangle)$ between the unknown state and its estimate. This defined by
\begin{equation}
I(|\psi\rangle, |\tilde\psi\rangle)=1-|\langle\tilde\psi|\psi\rangle|^2.
\end{equation}
For infinitesimally close states, the infidelity agrees with the Bures metric \cite{Hubner}. In addition, the inverse of the infidelity can be identified with the sample size required to reach a prescribed accuracy \cite{Mahler}. These two important properties of the infidelity motivate its use. However, it has been shown that states with a small infidelity might lead to very different physical properties \cite{Benedetti,Bina,Mandarino}. Consequently, other accuracy metrics have also been explored, such as, for instance, weighted mean-square error.

We generate a set  $\Omega_d=\{ |\psi_j \rangle \}$ of $m$ unknown pure states, which are identically, uniformly, and independently distributed. The transition probabilities toward the states of each basis are obtained by projecting each member of an ensemble of $N$ identically prepared copies of the unknown state. The transition probabilities are then estimated as
\begin{equation}
| \langle \psi_j | \varphi \rangle |^2 \approx  \frac{n_\varphi}{N},
\end{equation}
where $| \varphi \rangle$ is an element of one of the three bases, $n_\varphi$ is the number of outcomes in the direction of the state $| \varphi \rangle $, and $N$ the sum of the outcomes of all projections onto the elements of the basis. A total ensemble size of $3N$ is then employed as resource to obtain an estimate $|\tilde\psi\rangle$.

Due to the inherent randomness of the measurement process, the estimation of each state $|\psi_j \rangle$ is simulated $n$ times. This leads to $n$ different estimates $|\tilde\psi_j^{(i)} \rangle$ (with $i=1,\dots,20$). Thereafter, we average the infidelity over the $n$ estimates, that is,
\begin{equation}
I(|\psi_j \rangle)= n^{-1}\sum_{i=1}^{n} I(|\psi_j\rangle, |\tilde\psi_j^{(i)}\rangle).
\end{equation}
Finally, we calculate the mean of $I(|\psi_j \rangle)$ onto the set of unknown states, that is,
\begin{equation}
\bar{I} = m^{-1} \sum_{j=1}^{m} I(| \psi_j \rangle).
\end{equation}
In all numerical simulations, the bases ${\cal B}'_1$ and ${\cal B}'_3$ are fixed for all states in $\Omega_d$.

Figure\thinspace\ref{Figure1} displays, in logarithmic scale for both axes, mean and median of $I(|\psi_j\rangle)$ as a function of the total ensemble size $3N$ used in the estimation process for a single quantum system with dimension $d=4$ and for four randomly chosen unknown states in $\Omega_4$ (from top to bottom). The left column compares the mean infidelity generated by the 3BB-QT method (solid red dots) and the IC5BB-QT method \cite{CI5BB} (solid blue squares), a variation of the 5BB-QT method that achieves a higher accuracy. For the state at the first row of Fig.\thinspace\ref{Figure1} we see that both methods generate nearly indistinguishable results. The largest difference in the estimation accuracy is depicted at the third row of Fig.\thinspace\ref{Figure1}, where the IC5BB-QT method delivers and accuracy that is almost one order of magnitude better. The resting two states, second and fourth rows in Fig.\thinspace\ref{Figure1}, exhibit a small difference in the estimation accuracies for both methods. A similar behavior is exhibited by the median of the infidelity, which is depicted at the right column in Fig.\thinspace\ref{Figure1}. Furthermore, mean and median of $I(|\psi_j\rangle)$ reach very close values in the case of both methods. This is an indication that both methods do not exhibit outliers in the infidelity distributions.

\begin{figure}[t!]
\centering
\subfloat[Mean infidelity]{
    	\includegraphics[width=0.45\textwidth]{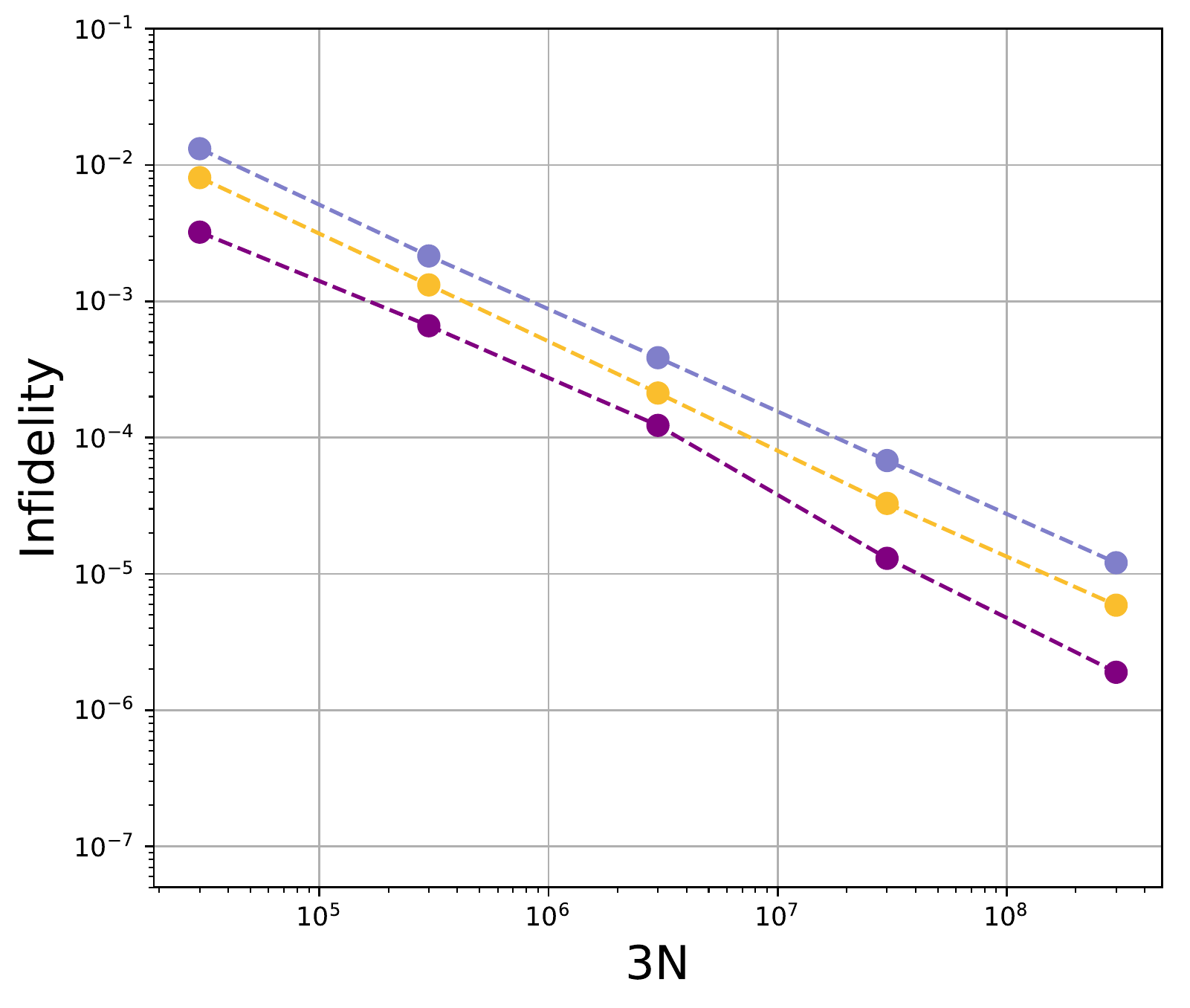}
    	\label{Fig2-1}}
\quad
\subfloat[Median infidelity]{
    	\includegraphics[width=0.45\textwidth]{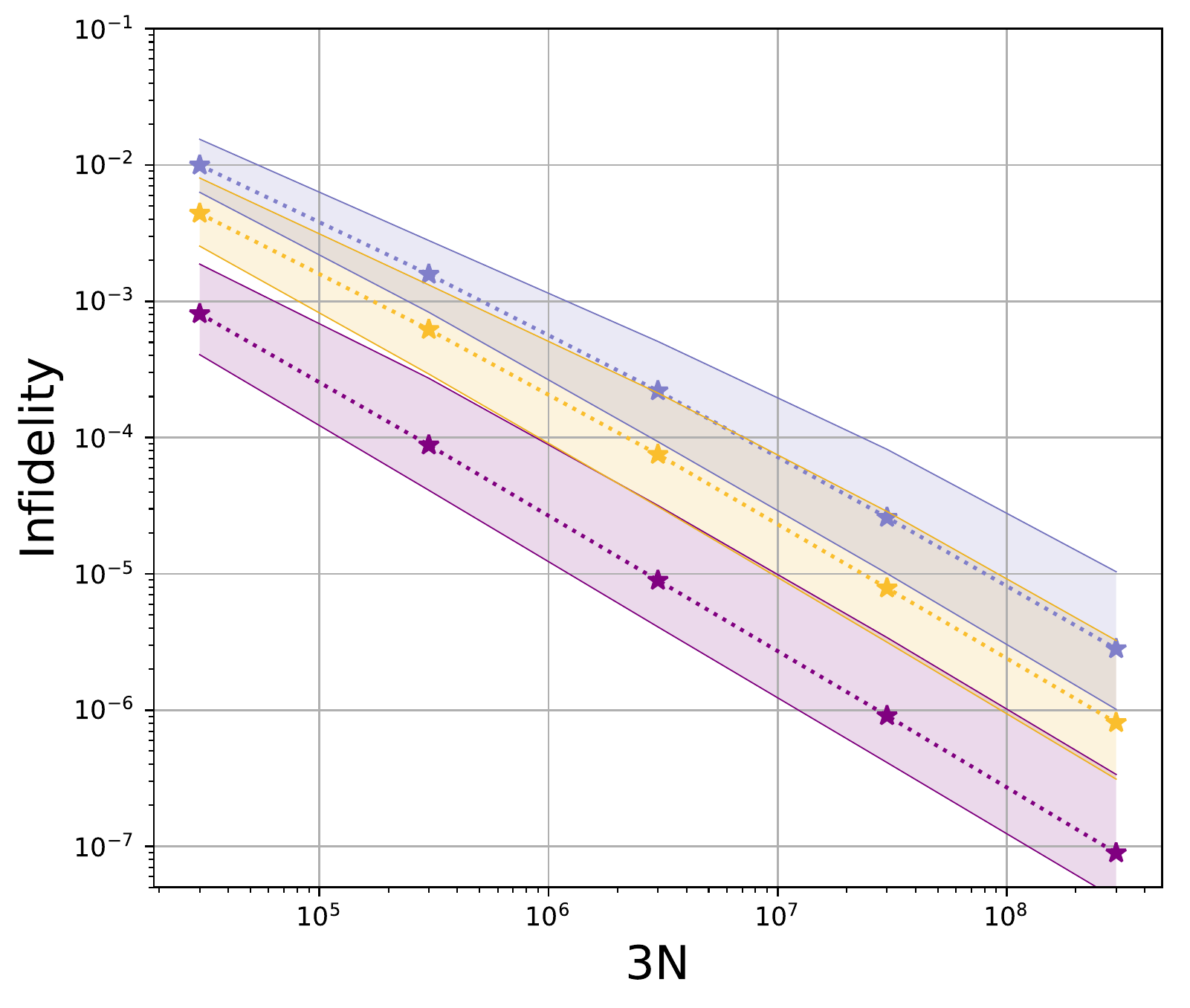}
   	 \label{Fig2-2}}
\caption{The mean (a) and median (b) of $I(|\psi_j\rangle)$ on $\Omega_d$ obtained via the 3BB-QT method as a function of total ensemble size $3N$ (in logarithmic scale for both axes) for dimensions $d$=4 (solid purple dots), $d$=8 (solid yellow dots), and $d$=12 (solid blue dots). Shaded areas represent the corresponding interquartile range.}
\label{Fig2}
\end{figure}

\begin{figure}[t!]
\centering
\subfloat[]{
    	\includegraphics[width=0.45\textwidth]{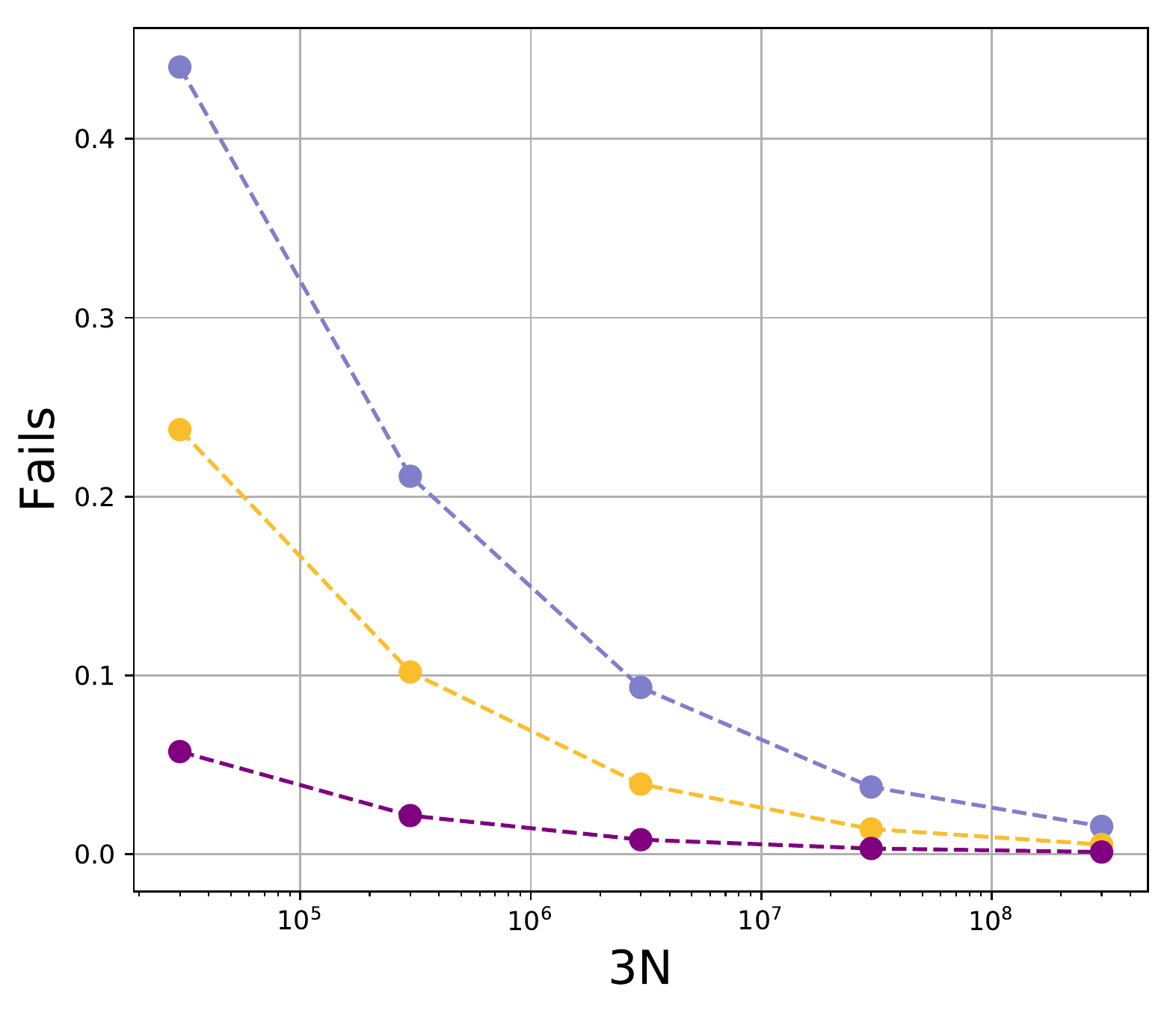}
    	\label{Fig3-1}}
\quad
\subfloat[]{
    	\includegraphics[width=0.45\textwidth]{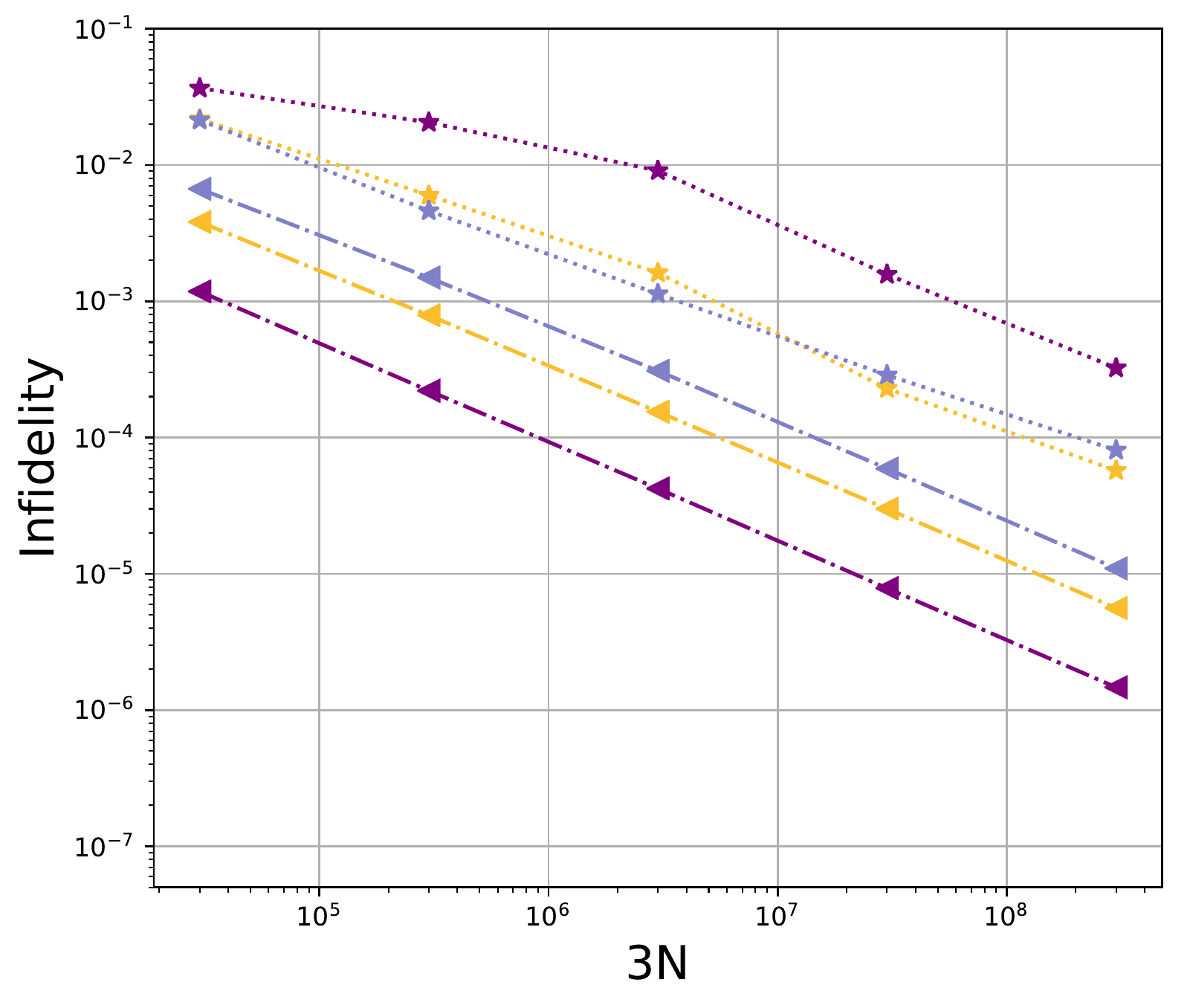}
   	 \label{Fig3-2}}
\caption{(a) Semi-logarithmic graph for the fraction of unknown states in the set $\Omega_f$ as a function of the ensemble size for dimension $d$=4 (solid purple dots), 8 (solid yellow dots), and 12 (solid blue dots). (b) Mean of $I(|\psi_j\rangle)$ on $\Omega_f$ as a function of total ensemble size $3N$ (in logarithmic scale for both axes) for dimension $d$=4 (solid red stars), 8 (solid green stars), and 12 (solid blue stars). (b) Mean of $I(|\psi_j\rangle)$ on the complement of $\Omega_f$ as a function of total ensemble size $3N$ for dimension $d$=4 (solid purple triangles), 8 (solid yellow triangles), and 12 (solid blue triangles).}
\label{Fig3}
\end{figure}


Figures \ref{Fig2-1} and \ref{Fig2-2} show the mean and median, respectively, of $I(|\psi_j\rangle)$ on $\Omega_d$ generated by the 3BB-QT method as a function of the total ensemble size $3N$ for $d$=4, 8, and 12, from bottom to top. A comparison between both figures reveals a median infidelity located below the mean infidelity for all values of ensemble size and dimension. In particular, the mean infidelity appears to be located at the upper border of the interquartile range. The gap between the median and mean infidelity seems to decrease with the increase of the dimension, and to increase with the increase of ensemble size. The existence of this noticeable gap between the median and the mean infidelity points out to the existence of states that are estimated with a low accuracy with respect to the value of the median.

The 3BB-QT method generates a set $\Omega_f$ of estimates that have a set $A_j$ with signs of the imaginary parts of $\Lambda_k$ that disagree with the unknown state. The states in $\Omega_f$ are characterized by an estimation accuracy lower than the mean of $I(|\psi_j\rangle)$ in $\Omega_d$. Figure\thinspace\ref{Fig3-1} shows the fraction of estimates in $\Omega_f$ with respect to a sample of $10^5$ unknown states as a function of the total ensemble size in dimension $d=$4, 8 and 12. As the figure indicates, higher dimensions exhibit larger fractions. This can be as large a 0.45 for $d=12$, that is, almost 45\% of all estimates are in $\Omega_f$. Also, the fraction is the largest for small ensemble sizes and rapidly decreases as the ensemble size increases. The average estimation accuracy of estimates in $\Omega_f$ is illustrated in Fig.\thinspace\ref{Fig3-2}, where the mean of the infidelity $I(|\psi_j\rangle)$ calculated on the states in the fraction is depicted as a function of the ensemble size in dimension $d=$4, 8 and 12. This set of estimates exhibits a mean which is one order of magnitud higher than the mean of the infidelity on $\Omega_d$. Fig.\thinspace\ref{Fig3-2} also shows the mean of the infidelity $I(|\psi_j\rangle)$ on the complement of $\Omega_f$. This is below the mean infidelity depicted in Fig.\thinspace\ref{Fig2} for small ensemble sizes. As the ensemble size increases the mean infidelity $I(|\psi_j\rangle)$ on the complement of $\Omega_f$ becomes very close to the mean infidelity on $\Omega$.

\section{Experimental state reconstruction of single-photon path-encoded qudits}

To experimentally test the method described in this work, we generate 8-dimensional qudit states encoded into the linear transverse momentum of single photons transmitted by diffractive apertures \cite{Neves05,Neves07,Glima08}. In this case, the dimension of the qudit state is determined by the number of paths available for the photon transmission over the aperture, which are typically addressed into spatial light modulators (SLMs) \cite{Glima09,MSolis01,Pimenta,Glima13,Rebon_2017}. Here, we define a multi-slit aperture into the used SLMs with eight parallel slits whose width is $3$ pixels wide and with $5$ pixels of separation, where each pixel is a square of $32\mu$m side length. Then, the state of the transmitted photons is given by \cite{Glima09}
\begin{equation}
|\psi\rangle=\frac{1}{\sqrt{N}}\sum_{l=-\frac{7}{2}}^{l=\frac{7}{2}} \sqrt{t_l}e^{i\phi_l}|l\rangle,
\end{equation}
where $|l\rangle$ represents the state of the photon transmitted by the $l$-th slit. $t_l$ and $\phi_l$ are the transmissivity and relative phase of slit $l$, respectively. $N$ is a normalization constant.

\begin{figure}[ht]
\includegraphics[width=0.45\textwidth]{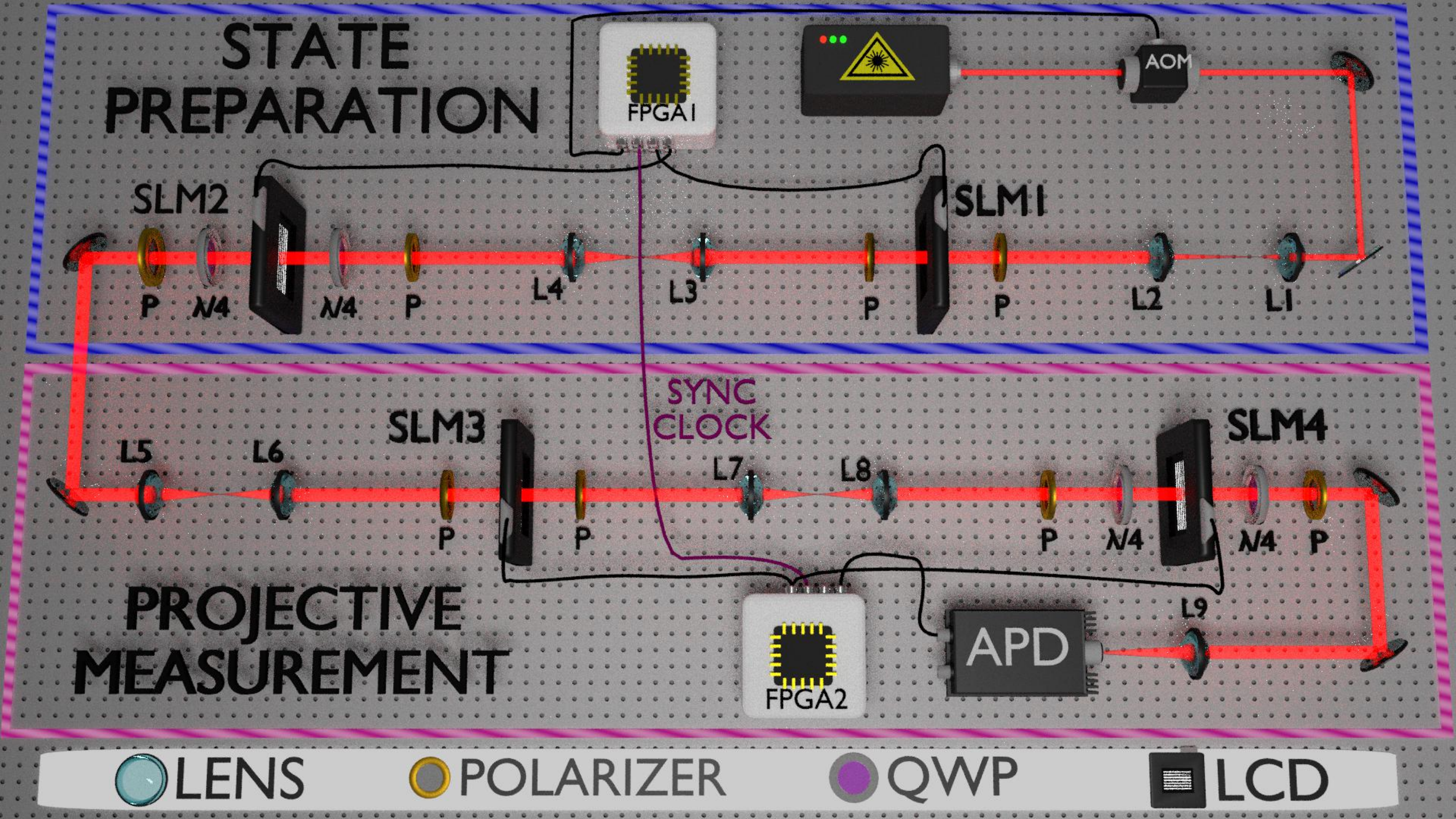}
\caption{The experimental setup has two parts, the state preparation (SP) stage and the projective measurement (PM) stage. The source consists of a CW laser, an AOM, and calibrated attenuators. The generated single-photons are sent through four transmissive SLMs placed in series, and with each LCD at the image plane of the previous one. SLM$1$ and SLM$2$ are configured to prepare the desired qudit state, which is then is projected onto any qudit state by means of SLM$3$, SLM$4$, and an APD. The setup is controlled and synchronized by FPGAs at each stage. See the main text for details.}
\label{Figure4}
\end{figure}

The setup consists of two parts, the state preparation (SP) stage and the projective measurement (PM) stage (see Fig.\thinspace\ref{Figure4}).  At the SP stage, the photon source is a continuous-wave (CW) laser, operating at $690n$m. It is combined with an acousto-optical modulator (AOM) to generate $40n$s wide pulses. Then, optical attenuators (not shown in the Fig.\thinspace\ref{Figure4}) placed at the output of AOM are used to create weak coherent states. The attenuators are calibrated to set the average number of photons per pulse to $\mu=0.9$. In this case, the probability of having pulses containing at least one photon is $P(\mu=0.9|n\geq 1)\approx 59.3\%$. Most of the non-null pulses contain only one photon and represent $61.7\%$ of the experimental runs. This type of light source is typically used in quantum information science since it can be seen as a good approximation to a non-deterministic source of single photons \cite{Gisin_2002,Lo_2014,Diamanti_2016,Xu_2019,Pirandola_2019}.

The generated photons are then sent through a first pair of transmissive SLMs (SLM1, SLM2).  Each SLM is composed of two polarizers, two quarter-wave plates (QWP), and a liquid crystal display (LCD). By proper configuring the polarizing optics, we set SLM1 and SLM2 to work in amplitude-only and phase-only modulation, respectively \cite{moreno}. SLM2 is located on the image plane of SLM$1$ via a $4f$ system with no magnification. An amplitude (phase) mask with eight slits is addressed on SLM1 (SLM2) with the gray levels of each pixel appropriately set to generate the desired initial state.

In order to measure over the $3d$ projective measurements required by the $3$BB-QT method, we employ a second pair of SLMs (SLM$3$, SLM$4$) and a  pointlike avalanche photodetector (APD). SLM$3$ and SLM$4$ are also configured for amplitude-only and phase-only modulation, respectively. The 8-slits addressed onto these last SLMs have the pixels's grey level adjusted to implement the projections required by the method. The projective measurement is completed with the APD positioned at the center of the transverse Fourier plane of the last lens $L9$. In this configuration, the single-photon detection rate is proportional to the overlap between the generated and post-selected states \cite{Glima13}.

To test our tomographic method, we considered the quantum state given by
\begin{equation}
|\psi_i\rangle=\frac{1}{\sqrt{8}}(|0\rangle-|1\rangle+|2\rangle-|3\rangle+|4\rangle-|5\rangle+|6\rangle-|7\rangle), \label{ExpQduit}
\end{equation}
and measure it using the $3$BB-QT bases (i.e., the diagonal basis and the bases of Eq.\thinspace(\ref{NEWB1'B2'})). During the measurement process, the SP and PM stages run automatically, synchronized and controlled by two field-programmable gate array (FPGA) electronic units at a rate of $10$Hz.  The FPGA$1$ is located at the SP stage and controls the first pair of SLMs and the AOM. In our case, the amplitude and phase masks deployed in the first pair of SLMs remain fixed during all experimental runs to generate the same initial state $|\psi_{i}\rangle$. In the PM stage, the FPGA$2$ controls the second pair of SLMs and records the number of counts detected by the APD. The synchronization between the two FPGA units allows us to project, for each coherent weak pulse, the prepared initial state into a different state. In order to minimize statistical fluctuations, the experimental setup automatically runs for $10$ hours.

\begin{figure}[t]
\centering
\includegraphics[width=0.45\textwidth]{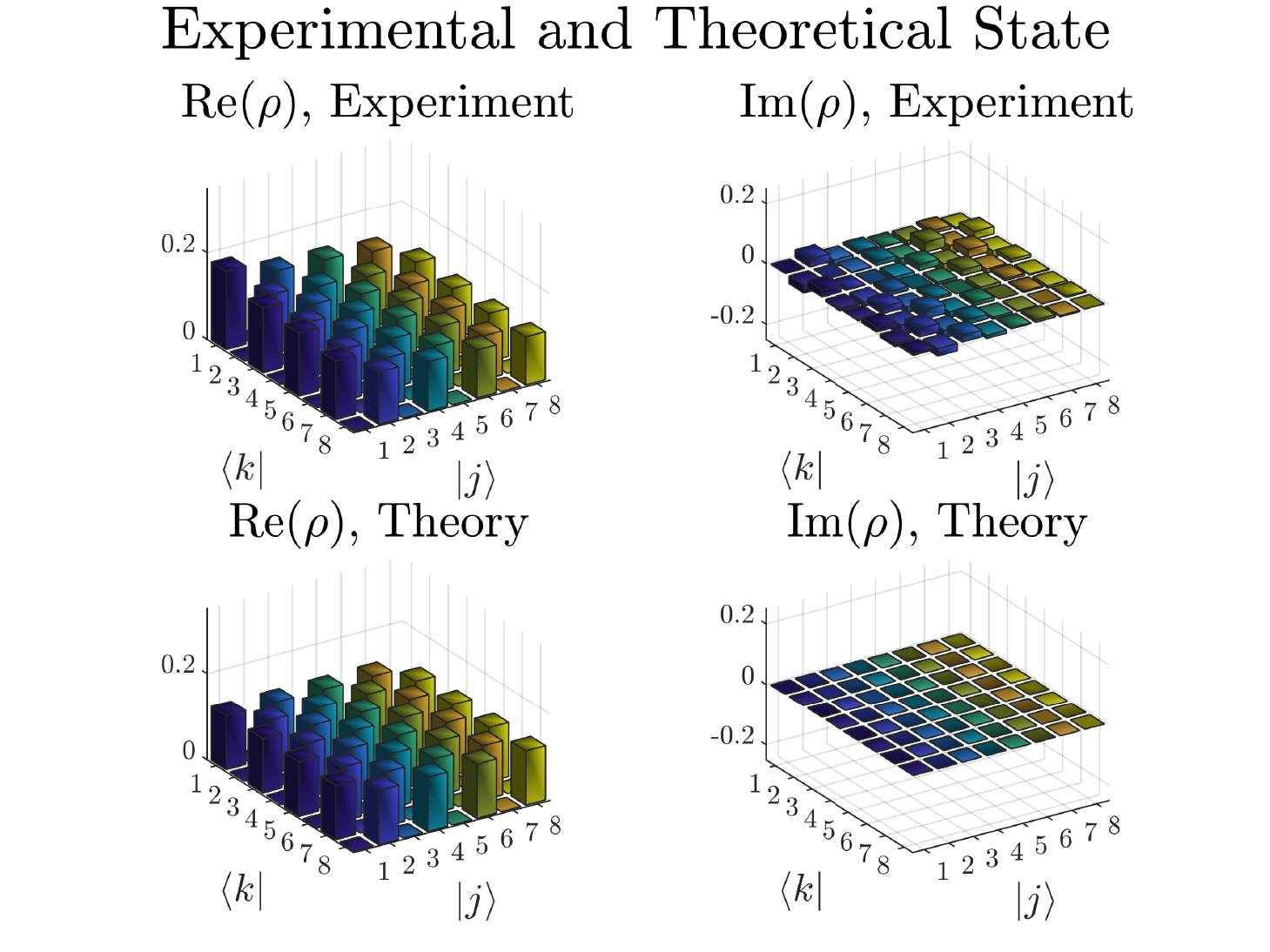}
\caption{Comparison between real and imaginary parts of the experimentally reconstructed state (upper row) and the target state of Eq. \ref{ExpQduit} (lower row).}
\label{Figure5}
\end{figure}

We calculate the probability distribution associated with the three $3$BB-QT bases from the recorded experimental data. With these probability distributions and applying the estimation method explained above, the infidelity between the estimated state and the target state $|\psi_{i}\rangle$ is only $0.0156$.  The state experimentally reconstructed is shown in Fig.\thinspace\ref{Figure5}. The upper row shows the real and imaginary part of the reconstructed density matrix, and the lower row shows the real and imaginary part of the target state. As one can see, the experimental results show the high-performance of the method for reconstructing high-dimensional quantum systems..

\begin{figure}[t]
\centering
\includegraphics[width=0.45\textwidth]{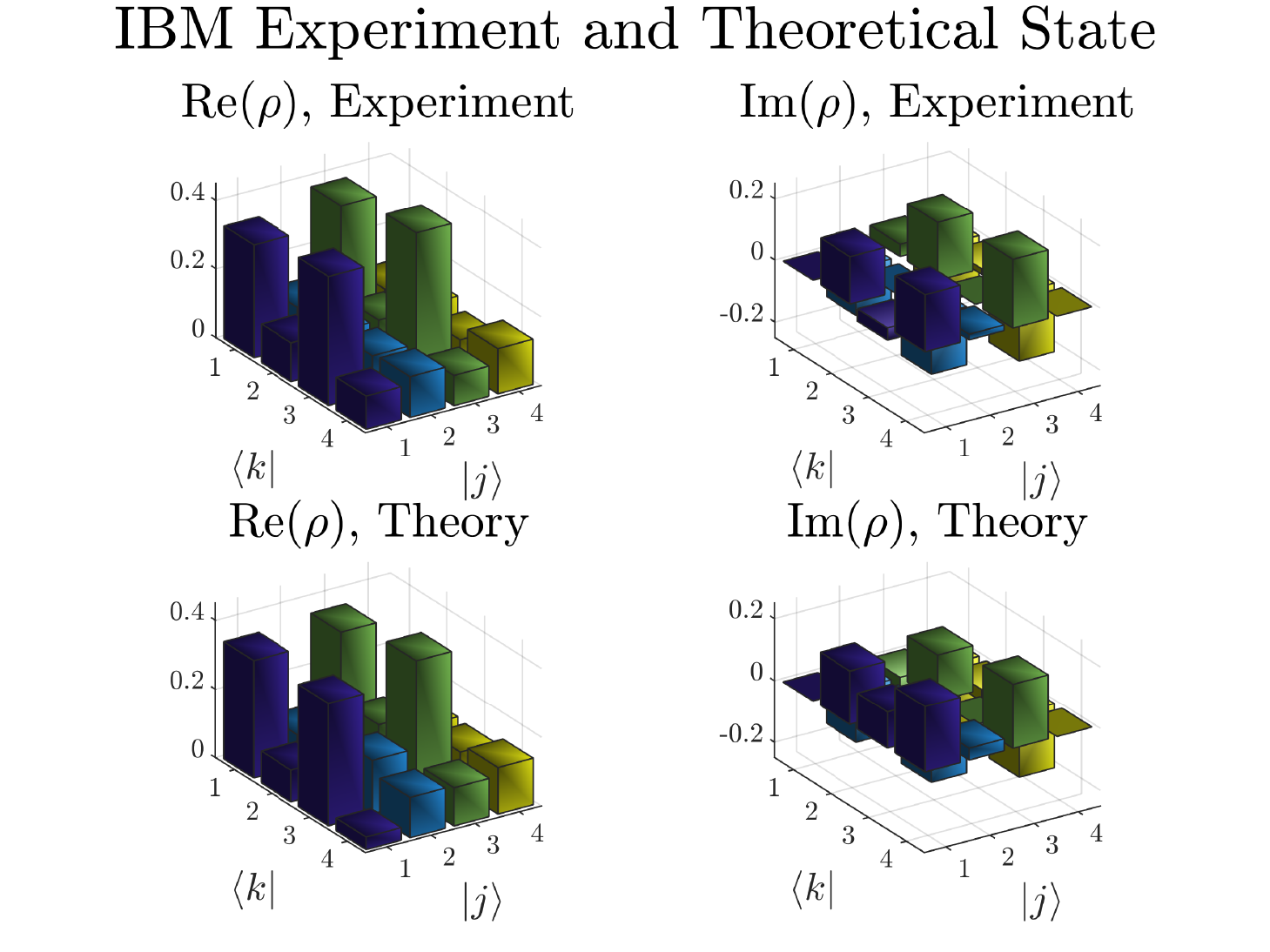}
\caption{Comparison between real and imaginary parts of the matrix coefficients of the experimentally reconstructed state (upper row) and the target state (lower row) for a two-qubit system on the IBM 5-qubit superconducting quantum processor.}
\label{Figure6}
\end{figure}

\section{Experimental validation in the IBM superconducting quantum processor}

We have also employed the 3BB-QT method to estimates states on the IBM Quantum Experience 5-qubit superconducting quantum processor ``ibmq-ourense". This quantum device has free access via its cloud service. In this quantum processor we can implement any unitary transformation acting on its qubits. Using Qiskit \cite{Qiskit}, an open-source development framework for working with quantum computers in Python, we can provide quantum circuits to the quantum processor, that are compiled into an equivalent circuit involving only the machine basis gates. The device only allows measurements on the computational base. Therefore, measuring on a non-computational basis $B$ requires applying $B^\dagger$ to a qubit and then measuring the resulting state on the computational basis. This procedure is repeated a fixed number of times to obtain the required statistics. This, however, introduces a noise in the inference of the probabilities that affects the performance of the 3BB-QT. This can be reduced by increasing the number of repetitions. On the other hand, there are systematics errors in the preparations of the gates. Due to the highly non-trivial preparation of the two-qubit CNOT gate, this has an error rate that is much higher than the one of local gates. This is the main source of error that affects the performance of the 3BB-QT method when implemented in the quantum processor.

To reduce the number of CNOT gates in our experiment, we have prepared the following separable state chosen at random
\begin{align}
	| \psi \rangle =& 0.5846|00\rangle + (0.157 + 0.295i)|01 \rangle \nonumber \\
	& + (0.608 + 0.200i)|10 \rangle + (0.062 + 0.362i)|11 \rangle,
\end{align}
and measured it using the diagonal base and the bases in Eq.\thinspace(\ref{NEWB1'B2'}). Each one of them corresponds to a different circuit, which we have measured using 8192 repetitions for each of them. This leads to a set of probabilities from which we estimate $\Lambda_k$ in Eq.\thinspace(\ref{Lambda_k}). Afterward, we  apply the estimation method. This procedure leads to an infidelity between the estimated state and the target state $|\psi\rangle$ of $0.035$ with an average error per gate of $0.005$. A comparison between the real an imaginary parts of the density matrix coefficients of estimate $|\tilde\psi\rangle\langle\tilde\psi|$ and target state $|\psi\rangle\langle\psi|$ is depicted in Fig.\thinspace(\ref{Figure6}), where a very good agreement can be observed.

\section{Conclusions}

In this article we have introduced a method to estimate pure quantum states of $d$-dimensional quantum systems. The method is based on three measurement bases in any dimension. Thereby, a total of $3d$ projective measurements are employed. In comparison, the 5BB-QT \cite{5B} method requires five observables or, equivalently, $5d$ projective measurements. The method employs $2d$ projective measurements to generate a finite set $\Omega$ with $2^{2d-1}$ pure states.  The estimate for the unknown state is given by the state in $\Omega$ with the highest likelihood. This is evaluated with the measurement results of the remaining $d$ projective measurements. We emphasize the fact that the likelihood is evaluated and not optimized, which contributes to reducing the computational cost of the method.

We have also studied the estimation accuracy achieved by the 3BB-QT method with the help of the infidelity as accuracy metric. We have shown by means of numerical experiments that the 3BB-QT method reaches an accuracy similar in median infidelity to the one provided by the 5BB-QT method. However, in the case of small ensemble sizes the mean infidelity provided by the 3BB-QT method is one order of magnitud higher than the one reached by the 5BB-QT method. The main reason is the fact that the unknown state and its estimate provided by the 3BB-QT method have different signatures of the coefficients $\Lambda_{k,\pm}$. This effect tends to vanish for higher ensemble sizes.

To experimentally prove the effectiveness of the 3BB-QT method, we performed two experiments on different and complementary high-dimensional quantum information platforms. First, we use a photonic platform that relies on the use of programmable spatial light modulators to prepare and measure single 8-dimensional quantum states encoded in the linear transverse momentum of single photons. Due to the high level of precision that can be obtained with the SLMs, the performance of the method can be properly studied and we demonstrate its efficiency by achieving an infidelity of only $1.5\%$ between the prepared and reconstructed state. At the second implementation of the protocol we used the 5-superconducting qubit IBM quantum processor ``ibmq-ourense", which served as a useful tool to study the performance of the method in a multipartite scenario. In this case, the observed infidelity was also very small giving $3.5\%$. These results together demonstrate the practicability of the method for the reconstruction of high-dimensional quantum states. Since the number of projective measurements required is only 3$d$, the 3BB-QT method is a powerful and interesting tool for the validation of novel quantum-based technologies.

\acknowledgments{This work was funded by the Millennium Institute for Research in Optics and by CONICYT FONDECYT Grants 1200859, 1190933, 3200779, and 1180558. L. Z. acknowledges support by CONICYT Grant 21181021.}

\end{document}